\documentclass[sigconf]{acmart}

\AtBeginDocument{%
  \providecommand\BibTeX{{%
    \normalfont B\kern-0.5em{\scshape i\kern-0.25em b}\kern-0.8em\TeX}}}

\begin{document}

\title{Microstructure-Empowered Stock Factor Extraction and Utilization}

\author{Xianfeng Jiao}
\authornote{Both authors contributed equally to this research.}
\email{jiaoxianfeng@stu.pku.edu.cn}
\affiliation{%
  \institution{Peking University}
  \country{China}
}

\author{Zizhong Li}
\authornotemark[1]
\email{zzoli@ucdavis.edu}
\affiliation{%
  \institution{University of California, Davis}
  \country{United States}
}

\author{Chang Xu}
\email{chanx@microsoft.com}
\affiliation{%
  \institution{Microsoft}
  \country{China}
}

\author{Yang Liu}
\email{yangliu2@microsoft.com}
\affiliation{%
  \institution{Microsoft}
  \country{China}
}

\author{Weiqing Liu}
\email{weiqing.liu@microsoft.com}
\affiliation{%
  \institution{Microsoft Research}
  \country{China}
}

\author{Jiang Bian}
\email{jiang.bian@microsoft.com}
\affiliation{%
  \institution{Microsoft Research}
  \country{China}
}

\def\xc#1{#1}
\def\xcc#1{\textcolor{black}{#1}}

\begin{abstract}
Stock investment has undergone a significant transformation driven by the development of computer technology, with high-frequency trading data emerging into prominence. Order flow data, the finest granularity of high-frequency data, made up of the order book and transaction data at the tick level, is crucial for market microstructure analysis as it provides traders with valuable insights to make informed decisions.
However, extracting and utilizing order flow data is challenging due to the large volume of data and the limitations of traditional factor mining techniques, which are designed for coarser-level stock data.
To address these challenges, we propose a novel framework to effectively extract important features from order flow data that can be applied at various temporal granularities.
Our method consists of a Context Encoder and an Informative Factor Extractor. The Context Encoder learns an embedding for the current order flow data segment's context by considering both the expected and actual market state. Meanwhile, the Informative Factor Extractor uses unsupervised learning methods to select such important signals that are most distinct from the majority within the given context. 
After that, {\xc{the relevant factors extracted from these signals are utilized in downstream tasks.
{\xcc{In empirical studies, we verify our method on an entire year of stock order flow data across diverse scenarios, 
thus providing a wider range of potential applications in comparison to existing tick-level approaches that have only been evaluated on small datasets spanning a few days of stock data.}}
We demonstrate that our method extracts superior factors from order flow data, enabling significant improvement for stock trend prediction and order execution tasks at the second and minute level.} 
\end{abstract}



\maketitle

\def\xc#1{#1}

\section{INTRODUCTION}
{\xc{The stock market holds a pivotal position in the financial industry, with quantitative analysis of stocks emerging as a highly significant research field within the realm of FinTech.
In recent years, there have been remarkable advancements in stock investment facilitated by rapid progress in computer technology and the availability of high-frequency data \cite{thakkar2021fusion,cavalcante2016computational,nazareth2023financial,bousbaa2023financial}. 
This abundance of high-frequency data, which is updated at a remarkably fast rate, ranging from every second to milliseconds, has opened up avenues for sophisticated analysis of financial markets. Consequently, it enables a deeper understanding of market microstructure and dynamics, thereby offering more accurate and real-time market intelligence.
}}

{\xc{In high-frequency data, \textit{order flow data} is the most detailed and raw data available for market microstructure analysis.}
More specifically, it is a chronological sequence of orders {\xc{at the tick level}} submitted to trading systems, including information such as the order type, price, volume, and the time of orders.
By recording real-time trading records, order flow data allows for the capture of subtle market changes such as trader sentiment and strategy that may be difficult to detect using market trading data at a coarser {\xc{temporal }}granularity.
For example, {\xc{as shown in Figure \ref{fig1}, order flow data can encompass the process of traders' ask and bid positions, capturing their order placement and transaction activities. It can dynamically depict the contrast of forces between buyers and sellers in real-time, providing a more immediate and detailed view of market dynamics. 
Specifically, the volume of sell orders is significantly higher than that of buy orders, indicating a large sell pressure. This imbalance in trading activity suggests a potential future drop in price, as larger selling pressure often leads to downward price movements.
Order flow data is in contrast to widely used candlestick charts, which can only represent lagging trading results.}} 
Therefore, order flow data can provide valuable insights and more precise market condition for traders to make more informed and instant trading decisions.

\begin{figure}[t]
\centering
\includegraphics[width=0.5\textwidth]{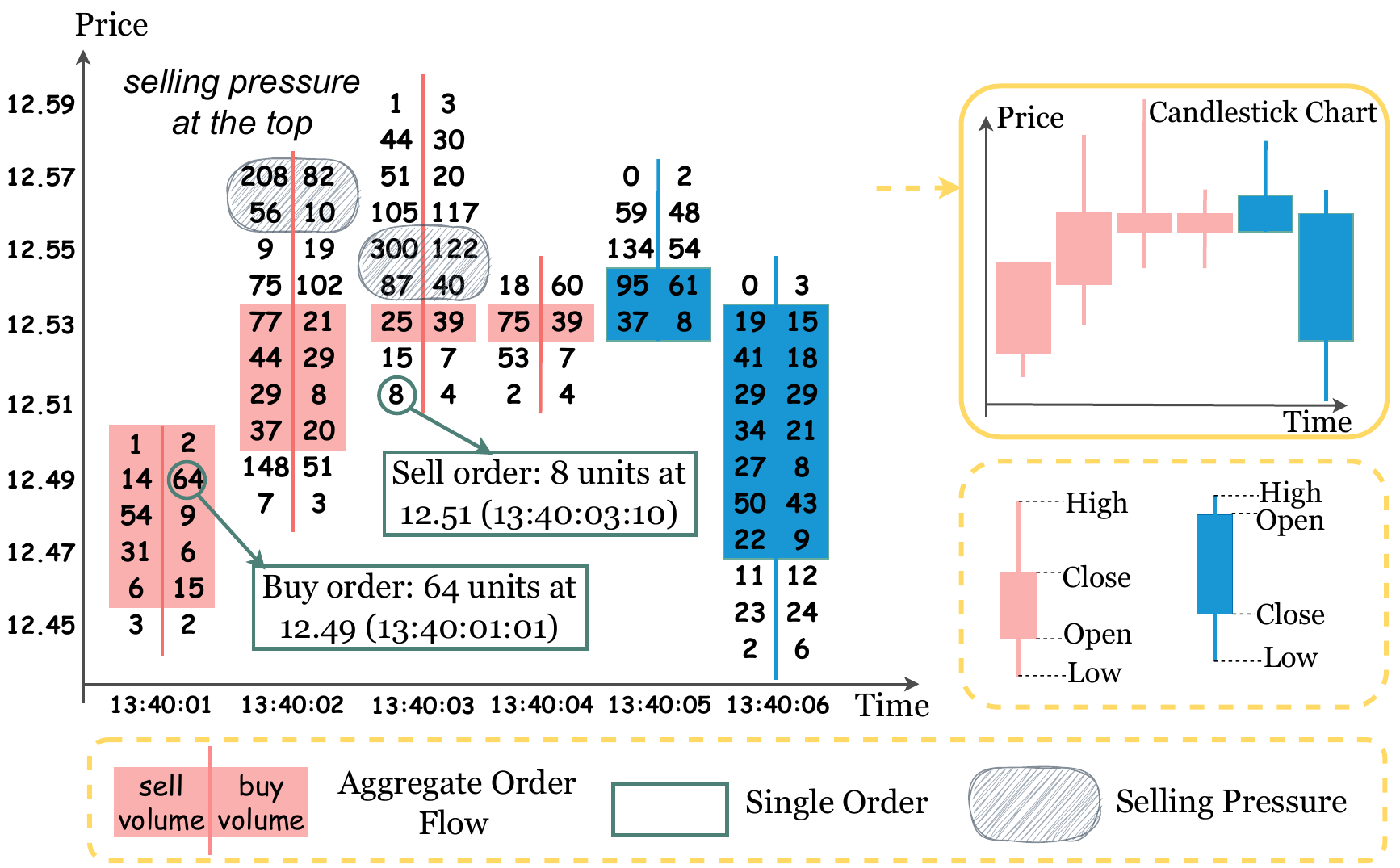} 
\caption{\xc{Aggregated order flow data (left) and candlestick chart (right) of the same stock. Numbers in colored blocks represent the aggregated volume of buy/sell orders at a specific prices over one-second interval. }}
\label{fig1}
\vspace{-5mm}  
\end{figure}

Despite the potential benefits, extracting and utilizing order flow data at granular levels remains challenging. For instance, the sheer volume of data with a single stock having upwards of 90,000 orders \footnote{Information obtained from the transactions of the Ping An Bank Co., Ltd. (000001.SZ) in 2020, see more in \href{https://finance.yahoo.com/quote/000001.SZ?p=000001.SZ&.tsrc=fin-srch}{\textit{Yahoo Finance}}.} traded in a single day. Since processing and analyzing such a huge amount of data is very difficult, traditional deep learning-based factor mining techniques typically operate at coarser time granularities, such as daily and hourly frequencies \cite{chen2019investment,wang2022adaptive,wangheterogeneous,feng2018enhancing,wang2021hierarchical,liang2021adaptive,li2021modeling}. This means that these methods may not be able to capture microstructural market features, such as the relative strength of buyers and sellers and trader sentiment and strategy, which are valuable in stock analysis.

Furthermore, extracting microstructural {\xc{factors from fine-grained}} data is only the {\xc{initial}} step. {\xc{Another key challenge lies in developing methods that offer adaptability across various downstream tasks and temporal granularities.}}
Previous research has examined \xc{leveraging} order flow data to {\xc{enhance downstream tasks}}, such as forecasting future {\xc{instantaneous }}stock movement trends \cite{liu2020multi,xu2022multi,mukherjee2023stock}, and {\xc{short-term }}order book reconstruction \cite{shi2021lob,shi2022state}. However, these studies primarily aim to extract microstructural information in the short term and {\xc{ may not directly address broader analysis over longer time periods. 
There is still a need for a method that provides greater adaptability across a wide range of downstream tasks.}}

In this paper, we aim to design {\xc{microstructure factor extraction and utilization}} methods \xcc{guided by the following principles}: (1) {\xcc{Microstructure market modeling:}} 
{\xcc{Focusing on modeling the microstructure market rather than adopting a macro perspective to extract meaningful signals of trading details from extensive order flow data.}}
(2) {\xcc{Generalizability across downstream tasks:}} Being generalizable across different types of downstream tasks, like return prediction and order executions at different coarser granularity (i.e. the daily-level and minute-level problems).
To handle the large volume of order flow level data, we divide the whole {\xc{order }}sequence into small segments, such as at the second level, and perform signal extraction on each segment {\xc{separately}}. 
Despite this segmentation, the number of segments for a broader time frame, such as daily {\xc{intervals, remains significantly challenging to handle.}}
Under these circumstances, we propose to {\xc{\textit{extract only the most important signals from these segments}}}. 
{\xc{Based on the principles of Information Theory \cite{shannon1948mathematical}, the segments that contain the most unique information are likely to be the most informative. 
In addition, to preserve the market state information of these segments, we encode both the expected and real market state as context.
Subsequently, we generate each segment's representation
{\xcc{and}}
employ unsupervised methods \cite{ruff2018deep} to select segments that exhibit the greatest dissimilarity from the majority given the context.
Then, the informative microstructure factors are extracted from these divided segments to serve different granularities of downstream tasks.}}
{\xcc{More specifically, our model comprises the following two essential components: Context Encoder and Factor Extractor.
For the Context Encoder, we utilize historical segments' order flow data to derive accumulated sell/buy orders as input.
Then, we employ separate generators for bid and ask orders, while incorporating shared information to generate future orders.
This enables us to generate a predicted order book for the next order segment, serving as a snapshot of the future market.
To assess the market state and its deviation from expectations, we introduce an \textit{expectation-reality comparison} method.
By comparing the predicted order book with the actual order book, we gain valuable insights into the market dynamics and the extent to which it aligns with expectations.
}}
{\xcc{The representation of the predicted and real order books for the next order segment are concatenated and used as the context information, indicating market state.}} 

{\xc{
Given the context information, the Factor Extractor utilizes transaction data, derived from order flow data, to extract vital features for the current segment. This data provides valuable insights into the dominant forces of both the buying and selling sides.
We design {\xc{\textit{conditional attention mechanism }}} to compress the information from transaction data while considering the current market state.
This is essential because the interpretation of the same transaction data stream can vary across different market states. For instance, in a scenario of significant upward price movement, a large volume of active sell orders often indicates profit-taking, whereas in a period of stable price sequences, it may suggest a bearish market sentiment leading to panic selling.
Subsequently, we use unsupervised learning to select the most distinctive signals from the extracted feature generated by Factor Extractor.
Finally, we extract feature factors based on specific tasks and concatenate the selected important trading signals from the chosen time segment range. These concatenated factors are then utilized in various coarse-grained downstream tasks.
}}

In empirical studies, we verify our method on real-world order flow data spanning up to one year across diverse scenarios, greatly exceeding the data range currently available for order flow work \cite{shi2021lob,xu2022multi,liu2020multi}. Also, it achieves superior performance compared to other microstructure-based factor mining methods in both daily return prediction and order execution downstream tasks. 

In summary, our main contributions are as follows:
\begin{itemize}
    \item[1)] We proposed a method that leverages a substantial amount of orders to model and extract vital trading details, thereby improving the analysis of stock trends. To the best of our knowledge, this study is one of the pioneering efforts in exploring the modeling of trading data from a microstructure market perspective instead of a macro view.
    \item[2)] We develop an innovative framework for extracting stock factor empowered by microstructure analysis, which utilizes historical order flow data to capture contextual information and employs a context-based factor extractor to select and extract significant stock factors from order flow segments.
    \item[3)] We demonstrate the effectiveness of our microstructure-based factor extractor in diverse downstream tasks with various temporal granularity by utilizing real-world data covering an entire year of order flow data, which provides a significantly broader range compared to previous order-level research.
\end{itemize}

\def\xc#1{#1}

\section{PRELIMINARIES}

In this section, we first review previous research in related areas, including stock factor extraction and utilization of microstructure market data. Then, we introduce order flow data and two types of data derived from order flow data: Limit Order Book (LOB) and transaction data, which are employed in our framework. For clarification and ease of understanding, Table \ref{tab:symbols} provides the list of notations used in our framework.

\begin{table}[t]
\caption{Notions Used in Our Model}
\centering
\begin{tabular}{c p{6.3cm}}
\hline
Notation & Definition \\
\hline
    $M$, $N$ & The {\xc{number}} of historical order flow segments and the number of one day's segments\\ \hline
    $t_0$, $\Delta t$ & The start time of each trading day and the {\xc{time length of one segment}}\\ 
    $p_x$, $w_x$ & The price and size of order $x$ \\  \hline
    $X$, $x$ & The order flow data and a single order \\ 
    $Xb_i$, $Xs_i$ & The order flow for buyers/sellers in a given time period \\  \hline
    $Z$, $z$ & Transaction data and a single transaction \\
    $O_t$, $\hat{O}_t$ & The order book at time $t$ and the predicted order book at time $t$ \\ 
    $\gamma(O_n, \hat{O}_n)$ & The measure of the difference between expected and actual order books for the n-th segment \\
    $\Delta Ob_i$, $\Delta Os_i$ & {\xc{Accumulated buy/sell orders in segment $i$}} \\ \hline
    ${p_b}^k_t$, ${p_s}^k_t$ & Price of buy/sell orders at price level $k$ \\
    ${v_b}^k_t$, ${v_s}^k_t$ & Volume of buy/sell orders at price level $k$\\ \hline

    $\oplus$ & The operation of combining two order books by summing volumes of each order book at corresponding price levels \\
    $\otimes$ & The {\xc{operation}} of generating new order books by previous order book and current order\\ \hline
    $G_b$, $G_s$ & The buy/sell generator used for generating accumulated buy/sell orders. \\
    $F_n^M$, $r_n^M$ & The context for the n-th segment given previous $M$ segments and the context representation \\ \hline
    $E_{trans}$ & The encoded feature sequence of transaction data \\
    $\widetilde{X}$ & The transaction feature $E_{trans}$ combined with context representation $r_n^M$ \\
    $Mat$ & The weighted masked matrix for Factor Extractor \\
    $F_{n}^{seg}$ & The extracted feature from $H$ heads for segment $n$ \\ \hline
    $\mu$ & 
    The proportion of order segment features used as factors based on the uniqueness indicator
    \\
    
\hline
\end{tabular}
\label{tab:symbols}
\end{table}

\subsection{Related Work}
Stock factor extraction involves identifying and analyzing patterns and trends in financial data to predict future stock performance, which can be applied in a variety of contexts, including portfolio construction, risk management, and trading strategy development. There are various methods and approaches to stock factor extraction, including statistical and machine learning techniques. Statistical methods \cite{rocsu2009dynamic,guilbaud2013optimal,avellaneda2008high,obizhaeva2013optimal} often rely on domain expertise and may not be able to generalize to dynamic stock markets. In recent research, deep neural networks have been proposed for learning stock factors. However, previous work \cite{chen2019investment,wang2022adaptive,wangheterogeneous,feng2018enhancing,wang2021hierarchical,liang2021adaptive,li2021modeling} has primarily focused on data at a coarse time granularity (e.g. daily) and has paid little attention to microstructure stock data. For example, these methods tend to model correlations between trading days rather than subtle patterns within intraday and order-level data. 

\begin{figure}[t]
\centering
\includegraphics[width=0.45\textwidth]{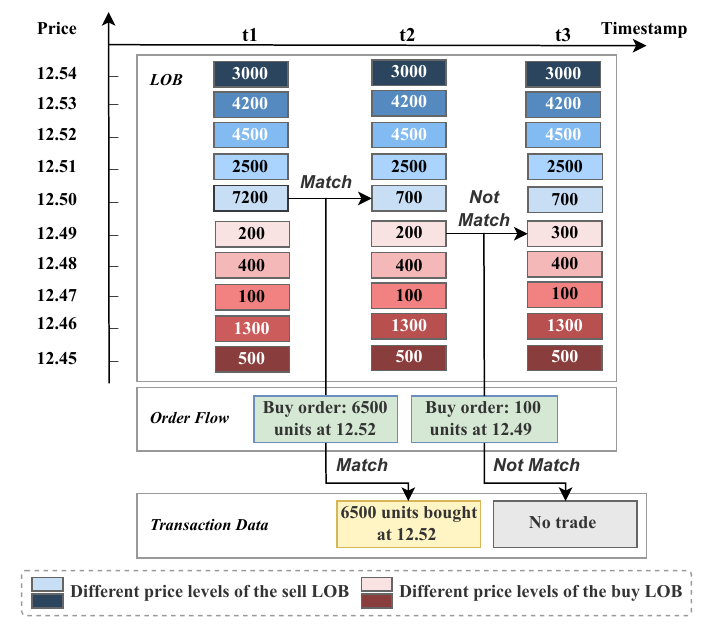} 
\caption{{\xcc{Illustration of dynamic updates of the Limit Order Book (LOB) based on the incoming order flow data. When a buy order matches with a sell order, a transaction is recorded and executed, leading to a change in the LOB. If no match is found, the LOB remains unchanged.}}}
\label{fig02}
 \vspace{-3mm}  
\end{figure}

The utilization of microstructure market data has been explored in a limited number of studies. Some of these works \cite{rocsu2009dynamic,obizhaeva2013optimal,avellaneda2008high,guilbaud2013optimal} have designed hand-crafted features to infer the instantaneous market state, such as the ratio of instant buy orders to buy orders\cite{lespagnol2018trading}. Other studies \cite{sirignano2019deep,liu2020multi,zhang2019deeplob,tsantekidis2017forecasting,magris2022bayesian} have applied deep learning techniques to tasks at the order or tick level, such as predicting instant stock prices\cite{liu2020multi,xu2022multi,ding2020hierarchical} or performing order-level market simulations \cite{mcgroarty2019high,wei2019model,moallemi2022reinforcement,schnaubelt2022deep,maglaras2022deep}. While these methods can capture instantaneous microstructure features, they may not be as effective as traditional stock factor extraction methods for tasks at daily or minutely time granularities.

{\color{red}
}


{\color{red}
}

\subsection{Order Flow Data}
Order flow data is a key component of the microstructure of a market. 
In this work, we adopt a definition of orders consistent with previous research \cite{gould2013limit,shi2021lob}.
Specifically, an order $x = (p_x, w_x, t_x)$ submitted at time $t_x$ with price $p_x$ and size $w_x > 0$ (respectively, $w_x < 0$) represents a commitment to sell (respectively, buy) up to $| w_x |$ units of the traded asset at a price no less than (respectively, no greater than) $p_x$.
\footnote{During preprocessing, we also convert market orders to their equivalent limit order form and filter out canceled orders.} 
The order flow is then defined as the ordered sequence of orders, i.e., the order flow $X = \{x_1, x_2, \dots, x_T\}$, where $t_{x_1} \leq t_{x_2} \leq \dots \leq t_{x_T}$.  
Order flow data represents the most granular and raw form of microstructure market data, and the limit order book data and transaction data derived from it are the types of data we specifically utilize in this work.
Figure \ref{fig02} illustrates the process of deriving transaction data and order book data from order flow data.


\noindent
\textbf{Transaction Data} 
Transaction data consists of records of completed trades in a market and typically includes information such as the price, size, and time of each trade. By analyzing transaction data, we can gain insights into various aspects of market microstructure, including the activity and intensity of the market and the behavior of large players participating in the market.
A single matching record at time $t_x$ can be described as 
$z = (p_z, w_z, t_z)$ with price $p_z$ and size $w_z > 0$ (respectively, $w_z < 0$) represents the active sell (respectively, buy) orders. 
Transaction data is a chronological sequence of orders, written as $Z = \{ z_1, z_2, \dots, z_L \}$.

\noindent
\textbf{Limit Order Book Data} 
The Limit Order Book (LOB) data provides a snapshot of the supply and demand for a given asset at a specific point in time. It can be used to infer the sentiment and intentions of market participants. The LOB consists of two main components: the bid side and the ask side, representing the highest prices at which buyers are willing to purchase the asset and the lowest prices at which sellers are willing to sell the asset, respectively. The LOB data typically includes the price, size, and time of each order placed.
Formally, we denote order books at time $t$ as $O_t=(Ob_t, Os_t)$, where $Ob_t$ and $Os_t$ represent buy order books and sell order books.
Buy order book at time $t$ is represented as 
$Ob_t = \{ ({p_b}^0_t, {v_b}^0_t), ({p_b}^1_t, {v_b}^1_t), \dots, ({p_b}^K_t, {v_b}^K_t) \}$, where ${p_b}^0_t > {p_b}^1_t > \dots > {p_b}^K_t$.
$({p_b}^k_t, {v_b}^k_t)$
represents that the volume of buy orders at price ${p_b}^k_t$ is ${v_b}^k_t$.
Similarly, we can get the sell order book $Os_t= \{ ({p_s}^0_t, {v_s}^0_t), ({p_s}^1_t, {v_s}^1_t), \dots, ({p_s}^K_t, {v_s}^K_t) \}$, 
where ${p_s}^0_t < {p_s}^1_t < \dots < {p_s}^K_t$.

To simplify notation and analysis, we assume that the order book is updated instantaneously when a new order $x_{t+1}$ is received. 
We use the operator $\otimes$ to denote the process of generating order book using previous order book and current order flow data, i.e., 
\begin{equation}\label{eq:merge}
O_{t+1}= O_t \otimes x_{t+1}.
\end{equation}
At $t=0$, the initial order book, represented by $O_0$, is empty, with a volume of zero at every price level.
(i) If the new order cannot be matched with the current orders on the order book, it will be added to the existing order book. For instance, if $w_{x_{t+1}} < 0$ (i.e., a buy order) and $p_{x_{t+1}} < p_s^0$ (i.e., cannot match), the order book is updated such that ${v_b}^k_{t+1} = {v_b}^k_{t} + |w_{x_{t+1}}|$ for the price level $k$ satisfying $p_{x_{t+1}} = {p_b}^k_t$.
(ii) If the new order can be matched with the current orders on the order book, it will consume them. 
For example, if $w_{x_{t+1}} < 0$ and $p_{x_{t+1}} \ge p_s^0$, the order book is updated such that ${v_s}^k_{t+1} = \max( {v_s}^k_{t} - (|w_{x_{t+1}}| - \sum_{i=0}^{k-1} {v_s}^i_t ), 0)$ for the price level $k$ satisfying $p_{x_{t+1}} = {p_s}^k_t$. 
(In cases of a partial match, we split the original order into two equivalent orders: a fully matched order and a remainder order that cannot be matched.)
At the same time, a corresponding transaction record is generated as $z = (p_z, w_z, t_z)$, where $w_z = w_x$. $p_z = (\sum_{i=0}^{k-1} {p_s}^i_t{v_s}^i_t + {p_s}^k_t{({v_s}^k_{t} - (|w_{x_{t+1}}| - \sum_{i=0}^{k-1} {v_s}^i_t ))} )/ |w_x| $ represents the actual transaction price.
(iii) If an order is canceled, the update corresponding to that order will be reversed and the order book will revert back to its previous state.
\def\xc#1{#1}
\def\xcc#1{#1}
\def\xccc#1{#1}
\def\xccc#1{\textcolor{magenta}{#1}}

\section{METHOD}
\begin{figure*}[t]
\centering
\includegraphics[width=0.98\textwidth]{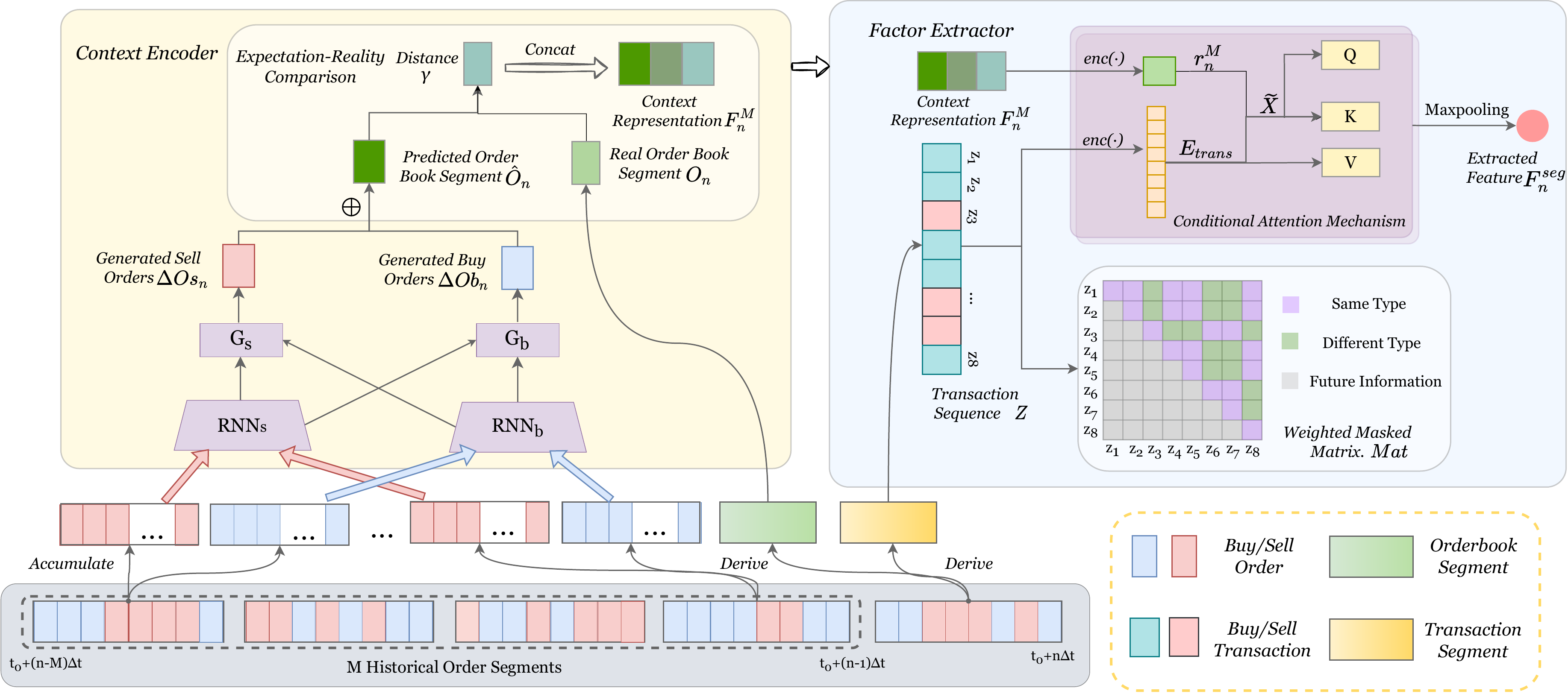} %
\caption{The framework of our microstructure-empowered stock factor extraction method. The Context Encoder (left) learns the context representation from both generated and real order book segments within historical order segments, further aiding in the extraction of order flow segment features. The Factor Extractor (right) utilizes content representation and transaction sequences to capture the ask and bid forces, extracting notable features from order flow segments.}
\label{fig2}
\vspace{-1em}
\end{figure*}

{\xc{
In this section, we will provide a detailed description of our approach for extracting microstructure factors from large-scale order flow data and demonstrate their utilization in downstream tasks. 
To handle the substantial volume of order flow-level data, we divide it into smaller segments and extract features from segments. 
In addition, we propose to consider the context of each segment during feature extraction.
In Section 3.1, we introduce how to encode historical information as market condition information by Context Encoder.
Section 3.2 presents our Factor Extractor module. 
To extract general factors applicable to diverse downstream tasks, we adopt unsupervised methods that do not rely on task-specific label information during feature extraction. 
Moreover, to apply the extracted information from the segments to tasks at coarser time granularity, 
we employ a selection mechanism that identifies the most influential subset of segments based on their deviations from the majority within the given context. 
We will now delve into a more comprehensive explanation of our proposed model.
}}

\subsection{Context Encoder}
We first divide one day's order flow into $N$ segments of small time span $\Delta t$, where the time span of the n-th segment is represented as $[t_0+(n-1)\Delta t, t_0+n\Delta t]$. We define $t_0$ as the start time of each trading day. To provide effective context information for the n-th segment, we leverage its previous M segments that have correlations in the information they cover.

Our approach is motivated by the concept that the order book, which captures the pending orders at a specific time, serves as a snapshot of the market, providing insights into traders' expectations regarding future market conditions \cite{pereira2020trading}.
Using historical data, we aim to predict the order book information for the n-th segment. 
By comparing the predicted order book with the actual one, we can infer the impact of historical information on the current state and use it as an important context for the n-th segment. 
For instance, if the predicted and actual order books are similar, we can conclude that the current state can be inferred from historical information, indicating the information from the n-th segment is less critical.
So we can encode market state information and the extent to which it aligns with expectations as context.
We will provide a detailed explanation of the components that make up the Context Encoder.

\smallskip
\noindent \textbf{Order Book Generator} 
In the stock market, there is often continuity or consistency in the trading behavior of the same direction  (buy/sell)\cite{barber2013behavior,jegadeesh2011momentum}. For instance, if a stock is consistently sold by sellers, it is likely to continue to be sold by sellers in the future. So, in order to model the behavior of buyers and sellers in financial markets, we train separate order book generators for buyers and sellers using historical order book and order flow information to predict the information of the {\xc{n-th segment's}} order book. Specifically, for the order flow segment $X_i$ of time period $[t_0+(i-1)\Delta t, t_0+i\Delta t]$, we first split it into buyer and seller order flows $Xb_i$ and $Xs_i$, respectively. 
We then calculate the accumulated volume of buy and sell orders at each price level in the $\Delta t$ time range, referred to as the accumulated {\xc{orders}} of the i-th segment, represented by $\Delta Ob_i = O_0 \otimes Xb_i$ and $\Delta Os_i = O_0 \otimes Xs_i$, where $O_0$ represents the original empty order book, and $\otimes$ is the operation of generating new order books as defined in Equation (\ref{eq:merge}). 

{\xc{
We encode historical M segments of accumulated buy/sell orders using sequential neural models such as RNN/LSTM.
For {\xcc{the segments $i = n-M, \dots, n-1$, we define the hidden state of the sell/buy order generator as:
\begin{equation}
    \begin{aligned}
            h_{b,i} = &RNN_b(h_{b,{i-1}}, \Delta Ob_{i})  \\
            h_{s,i} = &RNN_s(h_{s,{i-1}}, \Delta Os_{i}).
        \label{eq:Gs}
    \end{aligned}    
\end{equation}}}
We use $G_b$/$G_s$ to notate buy/sell orders generator. 
Since in real markets, the decision-making process of placing buy/sell orders often considers the actions of the opposing side, we incorporate information interaction in the buy/sell orders generator by {\xcc{sharing }}
their hidden states.
The generated accumulated buy/sell orders of segment n are given by
\begin{equation}
    \begin{aligned}
    \Delta Ob_n &= G_b(h_{b,n}, h_{s,n})\\
    \Delta Os_n &= G_s(h_{s,n}, h_{b,n}).
    \end{aligned}
\end{equation}
}}
Therefore, the predicted order book of the n-th segment is given by
\begin{equation}
 \hat{O}_n = O_{n-1} \oplus \Delta Ob_n \oplus \Delta Os_n,
\end{equation}
where $\oplus$ represents the operation of combining two order books by summing volumes of each order book at corresponding price levels to obtain a new order book.

\noindent \textbf{Expectation-Reality Comparison}
We define the difference between two order books as the total distance in volume at the same price levels. Here, we choose to use Euclidean distance. Therefore, for the n-th period, the measure of the difference between expected and actual order books is defined as:
\begin{equation}
\gamma(O_n, \hat{O}_n) = \sum_{k=1}^{K}||v_n^k - \hat{v}_n^k||_2,  
\end{equation}
where $K$ represents the number of price levels in the order book, and $v_n^k$ denotes the volume of orders at the $k$-th price level in the n-th period.
The value of $\gamma$ indicates the level of discrepancy between the expected and real order books. 
{\xcc{As our goal is to accurately model the behavior of buyers and sellers, we aim to optimize $G_b$ and $G_s$ by making the predicted order book as close as possible to the actual order book.}}
The training objective of buy and sell generators is defined as
\begin{equation}
\label{loss:ce}
\small
     L_{gen}=\frac{1}{N}\sum_{n=1}^{N}||\gamma(O_n, \hat{O}_n)||_2.
\end{equation}

\noindent \textbf{Context Representation}
We concatenate expected and actual order books, along with the measure of their difference, as context for the n-th segment given previous $M$ segments, denoted as $F_n^M = Concat(O_n, \hat{O}_n, \gamma(O_n, \hat{O}_n))$,
where $Concat$ is the concatenation operation. 

\subsection{Factor Extractor}
{\xcc{In this section, we introduce a method for encoding important information in the n-th segment by considering the context of $F_n^M$. Additionally, we aim to identify the most crucial segments among the N segments. Transaction data, which captures every trade that takes place, provides a detailed and precise overview of market activity and allows for the analysis of market microstructure. In our study, we utilize transaction information from the n-th segment to extract relevant features.}}

Transaction data consists of trades triggered by both active buy and active sell orders. An active buy order is placed by a trader who intends to purchase at the ask price, which is the price at which the seller is willing to sell the asset. This means that the trader is willing to pay a higher price for the asset. Conversely, an active sell order is an order filled at the bid price. 
Transaction data can provide insight into market sentiment and trader behavior. For example, a high ratio of active buy orders to active sell orders can indicate a bullish market sentiment, while a high ratio of active sell orders to active buy orders can indicate a bearish market sentiment. 

We simplify notation by omitting the segment index $n$ when it is clear from the context.
For the n-th transaction order sequence $Z=\{z_1, z_2, \dots, z_{L}\}$, we can get the active buy sequence $Zb=\{z_i\}, \forall v_{z_i} < 0$ and the active sell sequence $Zs=\{z_i\}, \forall v_{z_i} > 0$, where $L$ is the length of order sequence.
We then pass $Z$ through the feature encoding function $enc(\cdot)$, obtaining the transaction feature $E_{trans} = enc(Z) \in \mathbb{R}^{L \times d_e}$, where $d_e$ is the dimension of the feature encoding. 
This feature sequence is then input into a context-based multi-head attention module to extract a feature vector for the segment.

\noindent \textbf{Conditional Attention Mechanism}
We propose an improved version of the traditional multi-head attention mechanism that is conditioned on the context representation $r_n^M = enc(F_n^M)$, $r_n^M \in\mathbb{R}^{d_e}, $where $enc(\cdot)$ is a d-dimensional encoding function. The attention function maps a query and a set of key-value pairs to an output. Queries, keys, and values are represented by matrices $Q$, $K$, and $V$, respectively, defined as:
\begin{equation}\small
Q_i = \widetilde{X}W^{Q}_i,\
K_i = \widetilde{X}W^{K}_i,\
V_i = E_{trans} W^{V}_i,
\end{equation}
where $W^{Q}_i, W^{K}_i \in \mathbb{R}^{2d_e\times d_K}, W^{V}_i\in \mathbb{R}^{ d_e\times d_V}$ are weight matrices, $E_{trans}\in\mathbb{R}^{L\times d_e}$ is the feature sequence of the transaction data, and $\widetilde{X}\in \mathbb{R}^{L\times 2d_e}$ represents the feature combined with context, defined as $\widetilde{X}=Concat(E_{trans}, (\mathbf{1}_{d_e})^T r_n^M)$, where $\mathbf{1}_{d_e}$ is a $d_e$-dimensional vector whose elements are all 1.
By utilizing this approach, we incorporate the context information into the query $Q$ and key $K$, thereby enabling the consideration of context information during the attention computation.

\noindent \textbf{Weighted Masked Matrix} We design a weighted masked matrix $Mat$ to avoid future information leakage and distinguish the role of transactions of the same type from those of different types.
$Mat$ is formulated as
\begin{align}
    Mat_{i, j} = \begin{cases}0, &\text{when}\ t_{z_i} \leq t_{z_j}
 \\w,&\text{when}\ t_{z_i} > t_{z_j} \ \text{and} \ v_{z_i}v_{z_j} > 0
 \\1-w, &\text{when}\ t_{z_i} > t_{z_j}\  \text{and} \ v_{z_i}v_{z_j} < 0
\end{cases}
\end{align}
For the i-th head, the attention score matrix is defined as
\begin{equation}
    \mathbf{a_i}= softmax(\frac{Q_i K_i^T}{\sqrt{d_e}}\cdot Mat).
\end{equation}
{\xcc{In practice, we often observe that the influence between transactions of different types is more significant than that between transactions of the same type. To account for this, we assign a weight $w<0.5$ in the masked matrix, indicating a higher influence between different types of transactions.}}

{\xcc{To reduce the number of values for $L$ transactions, we apply a filtering process to retain only the most significant ones in each dimension. }}
The output of each conditional attention head is calculated as 
\begin{equation}
    h_i = maxpool(\mathbf{a_i} V_i),
\end{equation}
{\xcc{where $maxpool(\cdot)$ represents the max pooling operation, and $h_i \in \mathbb{R}^{d_{K}}$ is the resulting vector.}}
The final extracted feature from $H$ heads for segment $n$ is defined as 
\begin{equation}
    F_n^{seg} = Concat(h_1, \dots, h_H)W^H,
\end{equation}
where $W^H\in\mathbb{R}^{d_V\times d_e}$ {\xcc{is a weight matrix}}.

\begin{figure}[t]
\centering
\includegraphics[width=0.45\textwidth]{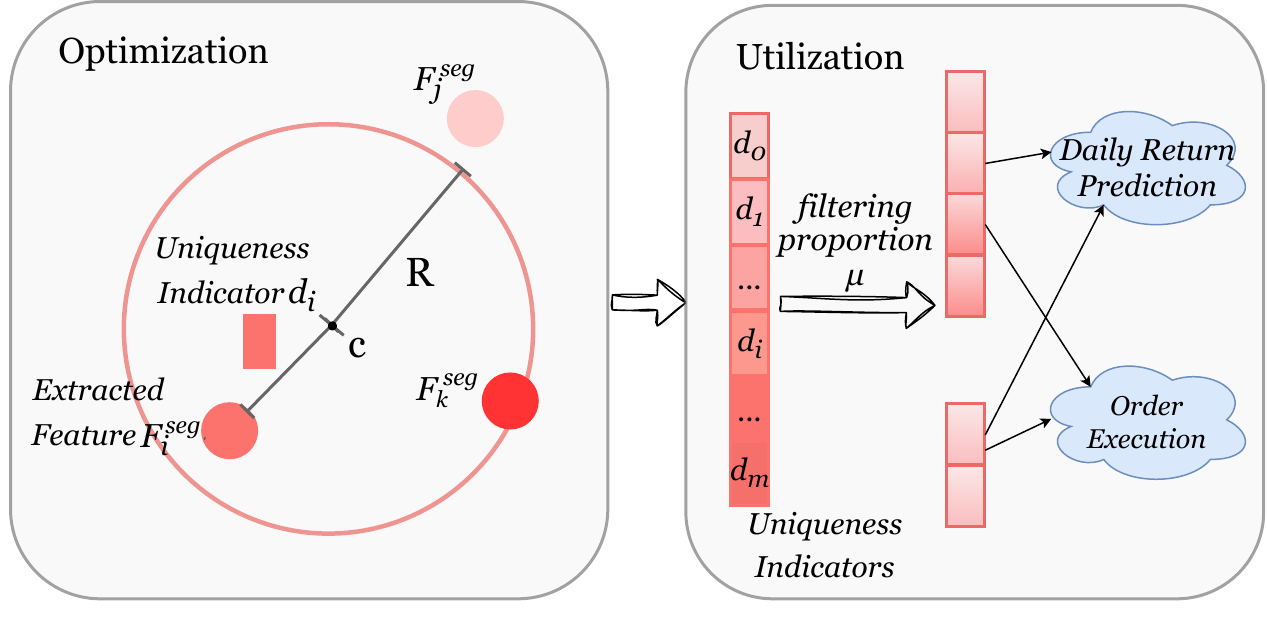} 
\caption{Left: Illustration of the optimization objective for selecting outliers among features extracted from our Factor Extractor. Right: Factor selection for different downstream tasks based on uniqueness indicators.}
\label{fig04}
 \vspace{-2mm}  
\end{figure}

\noindent \textbf{Optimization and Factor Utilization} 
In order to obtain general factors for various downstream tasks, we set the loss function as an 
unsupervised objective to discover the most distinct and valuable stock information.
We adopt DeepSVDD \cite{ruff2018deep}, one of the most effective unsupervised learning methods for identifying outliers in data streams. Figure \ref{fig04} illustrates the optimization and utilization processes in our method.
Specifically, we train a neural network as the kernel to map input space to a high-dimensional hypersphere (i.e. the output space). This hypersphere is defined by a radius $R > 0$ and a center $c$, allowing it to differentiate between normal data and outliers effectively. 
Given the context representation $F_i^{M}$ and the corresponding i-th transaction order sequence $Z_i$, the feature representation $F_i^{seg}$ can be generalized as:
\begin{equation}
    F_i^{seg} = f(Z_i|F^M_i;\theta)
\end{equation}
where $\theta$ is the parameters of our microstructure-empowered stock factor extractor.
Thus, given the sequence of input feature representations $\boldsymbol{F} = \{F_1^{seg}, F_2^{seg}, ..., F_N^{seg}\}$, the optimizing objective is defined as
\begin{equation}\label{objective}
    \min_{R, \theta} R^2+\frac{1}{\mu N}\sum_{i=1}^N\max\{0, ||F_i^{seg}-c||^2-R^2\}+\frac{\lambda}{2}||\theta||_2
\end{equation}
where hyperparameter $\mu \in (0,1]$ controls the trade-off between the size of hypersphere and violations of its boundary, $c$ is determined by random initialization. When applied to various downstream tasks, we utilize the parameter $\mu$ to control the filtering proportions of significant signals. Minimizing $R^2$ minimizes the volume of the hypersphere. The second term is a penalty term for points lying outside the sphere after being passed through the network (i.e. its distance to the center $||F_i^{seg}-c||$ is greater than radius $R$), and the last term is a weight decay regularizer on the network parameters $\theta$ with hyperparameter $\lambda > 0$.

Determined by the design of our optimization objectives, segments exhibiting the most significant differences in group properties are assigned a relatively larger distance $d_i=||F_i^{seg}-c||$ from the center $c$ of the hypersphere.
Hence, the distance $d_i$ can be used as the \textit{uniqueness indicator} for identifying and selecting segments that are highly distinct from large amounts of order flow data. Based on the \textit{uniqueness indicator}, we flexibly filter the number of signals (i.e., the selected features from order segment) by controlling the proportion coefficient $\mu$, thereby applying them to different granularity levels of downstream tasks.




\setlength{\tabcolsep}{1.5mm}{
\begin{table}[t]\label{tab:performance-drp}
    \centering
    \caption{Daily Return Prediction Performance of Our Method and Comparison Baselines on Stock Data.}
    \scalebox{0.95}{
    \resizebox{0.5\textwidth}{!}{
    \begin{tabular}{*{6}{c}}
      \toprule
      {\textbf{Method}} & \multicolumn{4}{c}{\textbf{Daily Return Prediction}} & \\
      \cmidrule(lr){2-6}
      & \textbf{IC$\uparrow$} & \textbf{Rank IC$\uparrow$} & \textbf{Rank IR$\uparrow$} &  \\
      \midrule
      Random Sample & 0.0118 & 0.0189 & 0.0534 \\
      Uniform Sample & 0.0212 & 0.0229 & 0.0684\\
      \midrule
      Price-Based Transaction Factor & 0.0292 & 0.0237 & 0.0730 & \\
      Volume-Based Transaction Factor & 0.0358 & 0.0322 & 0.0983 & \\
      Order Imbalance & 0.0350 & 0.0285 & 0.0871 & \\
      High-Freq LOB Feature & 0.0452 & 0.0412 & 0.1287 & \\
      Time-Sensitive Order Imbalance & 0.0364 & 0.0321 & 0.0984 &  \\
      
      \midrule
      Our Method (w/o CE,CAM) & 0.0561 & 0.0644 & 0.2050 \\
      Our Method (w/o CE) & 0.0529 & 0.0632 & 0.1992 \\
      Our Method (w/o Mat) & 0.0582 & 0.0670 &0.1908 \\
      Our Method & \textbf{0.0817} & \textbf{0.0787} & \textbf{0.2179} & \\
      \bottomrule
    \end{tabular}
    }
    }
    \label{tab:drp}
\end{table}}

\section{EXPERIMENT SETUPs}

In this section, we give a brief introduction of our experiment setting, the order flow dataset collected from the microstructure stock market, and two downstream tasks we use these order flow features to assist: \textit{Daily Return Prediction} and \textit{Order Execution (at the minute frequency)}. 
Generally, we first train the microstructure-empowered stock factor extractor that excels in selecting important trading signals. Then, we apply this trained upstream model to the practical implementation of downstream tasks with different levels of granularity.

\subsection{Dataset}
We evaluate our model on real-world stock data. Our dataset consists of two parts: candlestick (at the daily frequency) as baseline daily factor extractor dataset, and high-frequency (10-ms) order flow data as our intra-day micro-factor extractor dataset. Our dataset is collected from the 10 most active stocks of CSI300\footnote{CSI 300 is one of the major stock indices in the Chinese stock market, see more in \href{https://www.google.com/finance/quote/000300:SHA?sa=X&ved=2ahUKEwjx-aHjl9P8AhUDJkQIHcZoA3YQ3ecFegQIJBAY}{CSI 300 Index}.}. The dataset ranges from Jan.02,2020 to Dec.31,2020, and we split the dataset according to the ratio of 7/1/4 to form the train/validation/test dataset. 

For the daily candlesticks, we leverage 6 commonly used statistics attributes as daily features (i.e the highest price, the opening price, the lowest price, the closing price, the volume-weighted average price, and trading volume). For order flow data, we use a 4-second fixed time window and split approximately 4.56/0.67/2.64 million segments separately. We use price, size, and time (if applicable) as order flow features. All of the features are normalized by the Z-Score method to train our upstream and downstream models.

Our dataset consists of a whole year of ten representative stock order flow data instead of a few days of a handful of stocks, which are commonly used in the existing works based on the microstructure market\cite{shi2021lob,xu2022multi,liu2020multi}. Therefore, performing validation experiments or conducting practical applications on our dataset is able to provide a more reliable and comprehensive assessment on the focused microstructure stock market.

\subsection{Downstream Tasks}
In this work, we propose a novel upstream model for feature extraction and we showcase its effectiveness through two distinct downstream tasks: \textit{Daily Return Prediction} and \textit{Order Execution (at the minute frequency)}.

In the \textit{Daily Return Prediction} task, we aim to predict the daily return of a stock, which is $y=p_{T+2}/p_{T+1}-1$, where $p_t$ represents the close price of the stock at day $t$. For all of our comparison methods except the original daily prediction comparison method LSTM, we use order flow data to extract the intra-day factor as the micro supplementary information, concatenating with the daily factor generated by the LSTM baseline using candlesticks to make the daily prediction.

In the \textit{Order Execution} task, we set the problem definition following the problem definition in \cite{YuchenFang2021UniversalTF}, so our objective is to sell one unit of stock within a predetermined period and obtain the maximum profit. We use order flow factors to extract minute-frequency factors as additional input for the OPD teacher model. It will help the teacher model in training, and subsequently guide the training of the OPD student model.

\begin{table}[t]
    \vspace{-0.5cm}  
    \centering
    \caption{Order Execution Performance of Our Method and Comparison Baselines on Stock Data.}
    \small
    \begin{tabular}{*{5}{c}}
      \toprule
      {\textbf{Method}} & \multicolumn{2}{c}{\textbf{Stock}} & \\
      \cmidrule(lr){2-4}
      & \textbf{PA} & \textbf{GLR} & \\
      \midrule
      OPD & 0.58 & 0.94 \\
      \midrule
      High-Freq LOB Feature & 1.62 & 0.96 \\
      Price-Based Transaction Factor & 1.18 & 0.94 \\
      Volume-Based Transaction Factor & 1.03 & 1.00 \\
      Time-Sensitive Order Imbalance & 2.20 & 1.00\\
      Order Imbalance & 1.46 & 0.97 \\
      \midrule
      Our Method (w/o CE, CAM) & 2.23 & 1.03 \\
      Our Method (w/o CE) & 2.45 & 1.03 \\
      Our Method (w/o Mat) & 2.73 & 1.01 \\
      Our Method & \textbf{3.05} & \textbf{1.03} \\
      \bottomrule
    \end{tabular}
    
    \label{tab:oe}
\end{table}

\subsection{Baselines}

In our prediction tasks, we evaluate the performance of our model by comparing it with several competitive baselines. These are grouped into three distinctive groups:

\textbf{Group 1: Base Models}
This group comprises the foundational models upon which our prediction tasks are based.
\begin{itemize}
\item \textbf{LSTM} \cite{hochreiter1997long}: The original LSTM model uses only daily-frequency representations and no intra-day features.
\item \textbf{Oracle Policy Distillation(OPD)} \cite{YuchenFang2021UniversalTF}: OPD is a method for training deep reinforcement learning agents that aim to reduce execution costs for trading.
\end{itemize}

\textbf{Group 2: Feature Extraction Methods}
The second group includes various methods for extracting intra-day features from stock transaction records. These extracted features serve as auxiliary information to be fed into the base models to improve their predictive performance.
\begin{itemize}
\item \textbf{Random Sample}: Randomly extracted intra-day feature within single-day transaction record of stock.
\item \textbf{Uniform Sample}: Timed uniformly extracted intra-day feature within single-day transaction record of stock.
\item \textbf{High-Freq LOB Feature} \cite{mcgroarty2019high}: Intra-day feature vectors of a stock can be extracted directly from LOB, including time-insensitive feature vectors and time-sensitive feature vectors.
\item \textbf{Price-Based Transaction Factor}: Extracted intra-day feature based on the maximum and minimum trade price within a single-day transaction record of stock.
\item \textbf{Volume-Based Transaction Factor}: Extracted intra-day feature based on the maximum and minimum trade volume within a single-day transaction record of stock.
\item \textbf{Order Imbalance Factor} \cite{lespagnol2018trading}: A classical method to measure the imbalance of order by computing the log difference of the five best bids/ask level depth from the current depth of the best asks/bids.
\item \textbf{Time-Sensitive Order Imbalance Factor} \cite{chordia2004order,shen2015order}: Time-Sensitive order imbalance measures the difference between time-based limit ask and bid orders sequence using different level depth's volume, which is an important signal in market trading.
\end{itemize}

\textbf{Group 3: Our Proposed Method and its Variants}
This group consists of our novel feature extraction model and its variants. They also extract features to be inputted into the base models to verify their effectiveness.
\begin{itemize}
\item \textbf{Our Method (w/o CE, CAM)}: This variant removes the Context Encoder and Conditional Attention Mechanism modules from our proposed method.
\item \textbf{Our Method (w/o CE)}: This variant removes the Context Encoder and uses current real order book data as the context feature for the Conditional Attention Mechanism module.
\item \textbf{Our Method (w/o Mat)}: This variant removes the Weighted Masked Matrix from our proposed method.
\end{itemize}

The features extracted by the methods in Groups 2 and 3 are added to the base models of Group 1 to assess their effectiveness.

\subsection{Implementation Details}
In this subsection, we introduce the practical implementation settings used in our upstream model and downstream tasks.

\noindent
\textbf{Context Encoder}
For the \textit{Context Encoder}, we set $M=100$. The $RNN_s$ and $RNN_b$ are implemented as LSTM models, with hidden size of 64. $G_s$ and $G_b$ are implemented as MLP with hidden size 64. 

\noindent
\textbf{Factor Extractor}
For the \textit{Factor Extractor}, we use a 1-layer \textit{Conditional Attention Mechanism} with a hidden size of 16. The number of attention heads is set to 4. For the optimization objective expressed in Equation (\ref{objective}), we set $\mu$ to 0.02 for the ratio of anomaly data and set $\lambda$ to 0.1 for the coefficient of $L_2$ regularization. 

For both components, the time step $\Delta t$ is set to 4s, the learning rate is set to $1 \times 10^{-3}$, and the Adam optimizer is used. All computations were performed on a Nvidia A100 GPU.


\noindent 
\textbf{Daily Return Prediction Settings} 
To evaluate the performance of our prediction task, we use the top 2\% of trading/order sequences as representative intra-day micro supplementary information and select related factors (i.e. price, volume, time, factor) from these sequences. These sequences are then processed by an MLP layer to generate a daily micro-mining frequency representation.

We use the Information Coefficient (IC, also known as the Pearson Correlation Coefficient), Rank Information Coefficient (Rank IC, also known as the Spearman Correlation Coefficient), and Rank Information Ratio (Rank IR) to evaluate the accuracy of daily return prediction. In particular, Rank IR is a measure of the consistency and robustness of the predictions.
\begin{equation}
Rank IR = \frac{\text{Mean}(Rank IC)}{\text{Std}(Rank IC)}
\end{equation}


\noindent 
\textbf{Order Execution Settings}
To evaluate the performance of our execution task, we extract the mean, median, maximum, minimum, variance, and range of the order flow factors at a minute frequency as supplementary features for the OPD teacher model, which then guides the training of the OPD student model.

We use the Price Advantage (PA), and Gain-Loss Ratio (GLR) as in \cite{YuchenFang2021UniversalTF} and compared their performance. PA and GLR are represented by the following equations, where $\overline{P}^k_{strategy}$ is the corresponding average execution price that our strategy has achieved on order $k$. $\widetilde{P}^k$ is the average market price of the instrument on the specific trading day. $|D|$ is the size of the dataset.
\begin{small} 
\begin{equation}
    PA = \frac{10^4}{|D|}\sum_{k=1}^{|D|}(\frac{\overline{P}^k_{strategy}}{\widetilde{P}^k} - 1)
\end{equation}
\end{small}
\begin{small} 
\begin{equation}
    GLR = \frac{E[PA|PA>0]}{E[PA|PA<0]}
\end{equation}
\end{small}

\def\xcc#1{#1}

\section{RESULTS}
{\xcc{In this section, we demonstrate the validity of our model on real-world stock markets dataset. 
}}
In order to validate that the factors extracted using our proposed framework can provide valuable information for stock trend prediction and support tasks at different granularity levels, we compare our method with a range of advanced baselines for both \textit{daily return prediction} and \textit{order execution} tasks. Our method shows promising results in extracting the most informative trading moments of the day from microsecond-frequency trading data, and it can be used for both two downstream tasks with daily and minute frequencies. 
Moreover, we conduct a series of ablation experiments, proving that our designed modules, namely the \textit{Content Encoder}, \textit{Conditional Attention Mechanism} and \textit{Weighted Masked Matrix}, toward enhancing the overall performance of our method.
Furthermore, in order to gain a deeper understanding of the model's performance, we conducted a case study within a specific trading segment of individual stocks. Our observations revealed a strong alignment between the significant trading information segments extracted by our model and instances of imbalances within the order book, showing our model's ability to capture and reflect relevant financial phenomena.
\subsection{Comparison with Baselines} 
\noindent 
\textbf{Daily Return Prediction} The performance of our method and other baselines of the daily return prediction downstream task are shown in Table \ref{tab:drp}. Comparing the simple feature extraction methods (i.e., Random Sample and Uniform Sample) to other rule-based methods in Group2, we can observe that incorporating intra-day features selected through reasonable rules rather than random selection from stock transaction records can enhance the performance of the model in predicting daily return. Specifically, compared to Random Sample and Uniform Sample, the other rule-based baselines in Group2 exhibits a noticeable improvement in all three metrics, which means selecting inter-day features with certain distinct characteristics in stocks is indeed important and can assist the model in predicting stock trends. 

Building upon this foundation, our proposed model achieves even better performance compared to the best-performing rule-based baseline (i.e., High-Freq LOB Feature), consistently achieving the best results across all evaluation metrics. Specifically, our approach outshines the best performing baseline method by 81\% in IC, 91\% in Rank IC, and 69\% in Rank IR. These substantial improvements affirm the robustness and superior performance of our method in capturing the intricate dynamics of daily return predictions. In contrast to other baselines that rely on rigid rule-based factor extraction, our method demonstrates greater flexibility and adaptability to the ever-changing nature of stocks.


\noindent 
\textbf{Order Execution} Unlike daily return prediction, which operates at a daily frequency, order execution is performed at a more granular level, specifically in minutes. 
The performance of our method and other baselines for order execution is shown in Table \ref{tab:oe}. Compare to the original OPD model, incorporating all of the rule-based methods from Group2 to extract intra-day features and integrating them into OPD leads to an enhancement in model performance. 

Furthermore, compare to Time-Sensitive Order Imbalance baseline, the rule-based model which shows the best performance in extracting intra-day features to support OPD, our proposed machine learning-based method exhibits an improvement of 38.64\% in PA, and an improvement 3\% in GLR, demonstrating our method's effectiveness in optimizing the order execution task.

\def\xcc#1{#1}

\subsection{Ablation Study}


Table \ref{tab:drp} and \ref{tab:oe} present the results of the ablation study on the proposed method. Our method(w/o CE, CAM) refers to removing the Context Encoder and Conditional Attention Mechanism modules. Our method(w/o CE) refers to removing the Context Encoder and using current real order book data as the context feature for the Conditional Attention Mechanism module. Our method(w/o Mat) refers to removing the Weighted Masked Matrix.

The results indicate that our method(w/o CE, CAM) outperforms all other baselines on the two downstream tasks, however it falls short of the performance of our method(w/o CE). Furthermore, our method demonstrates superior performance over these two variants. This suggests that historical order books can provide effective contextual information for modeling transaction data segments and supports the effectiveness of our Context Encoder in extracting market expectation information on the current state. Additionally, the performance of our method(w/o Mat) was also evaluated and resulted in a decline in performance. This further verifies that the Weighted Masked Matrix module is reasonable for modeling the dynamic interactions between buyers and sellers in transaction data and brings performance improvement.
\vspace{-1.8mm} 


\begin{figure}[t]
\centering
\includegraphics[width=0.5\textwidth]{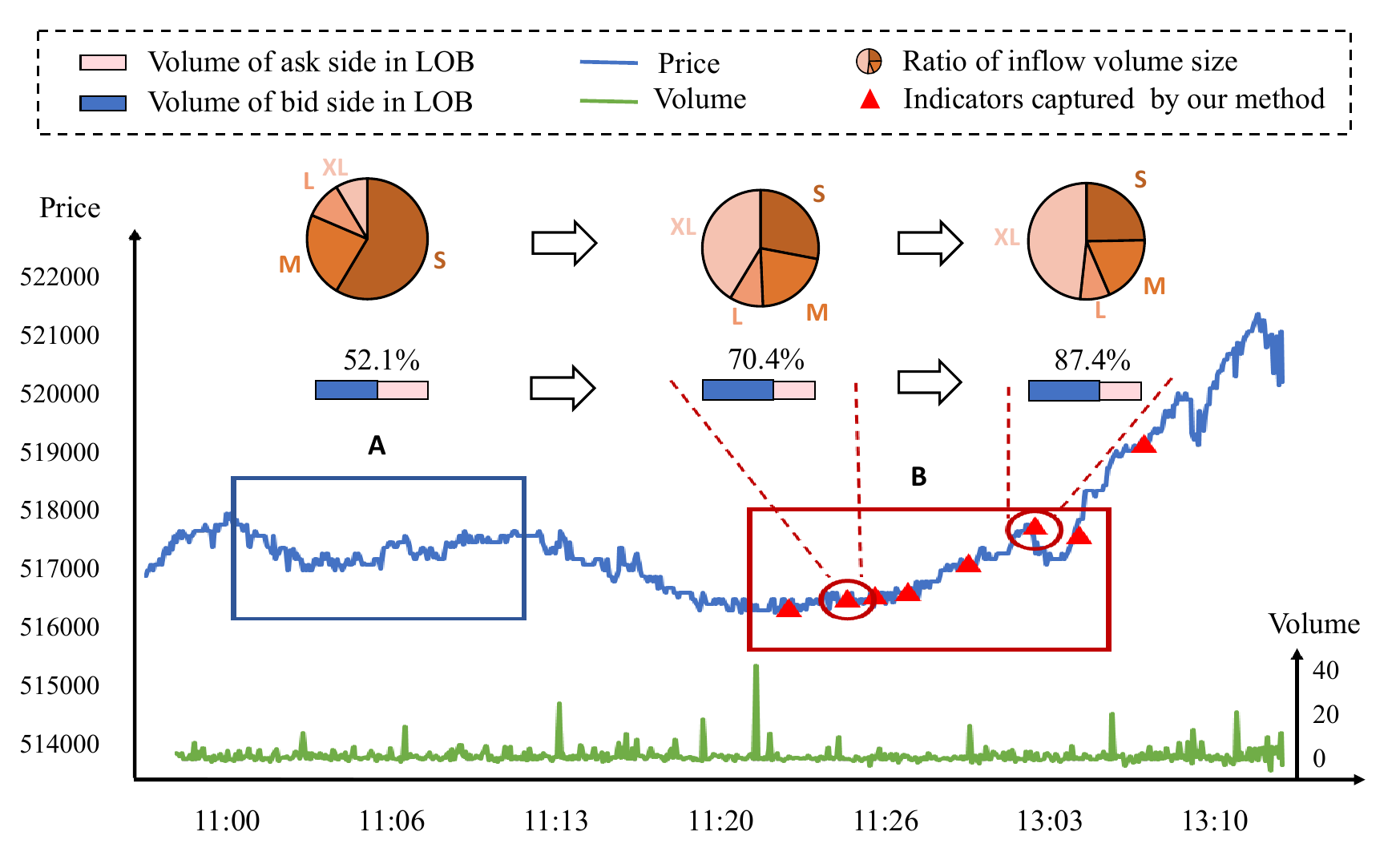} 
\caption{Case study of a randomly selected test case\protect\footnotemark, visualizing uniqueness indicators alongside the price-volume trend and the changes in the composition of market order sizes and directions. Our method identifies key time points of institutional involvement and significant imbalances in the order book.}
\label{fig:case}
\vspace{-7mm}  
\end{figure}
\footnotetext{On November 30, 2020, we selected stock 000568.SZ from the China stock market as our target for analysis.}

\subsection{Case Study}
{\xcc{To get a deeper understanding of how our method works, we randomly sample a trading day and visualized the uniqueness indicators captured by our approach.
 }
In Figure \ref{fig:case}, "Volume of bid side in LOB" refers to the volume of buy orders at the top 5 levels of the order book. "Ratio of inflow volume size" \footnote{ The division rules of larger(XL/L) order and smaller(M/S) order can be seen in \href{https://support.futunn.com/en/topic498/?url=202010641&skintype=1&clientlang=2&clienttype=12&user_id_type=1&clientver=12.46.9208&is_visitor=0&user_id=33847042&main_broker=WwogIDEwMDEKXQ==&user_main_broker=WwogIDEKXQ==}{Ratio of inflow volume size}.} refers to the proportion of larger and smaller orders. 

{\xcc{The price trend in the market is influenced by the ongoing competition between long and short positions, and identifying the turning points in this competition is crucial. }}
As shown in Figure \ref{fig:case}, our method identifies a concentrated accumulation of uniqueness indicators (i.e., red triangles) in interval B (11:18-13:03). 
We can discern that the B interval aligns with the turning point of the trend, which demonstrates the effectiveness of our method in capturing stock trend information through uniqueness indicators.

From a deeper perspective of stock trading, by comparing the order flow data in interval B to the fluctuating interval A (11:00-11:10), we observe a significant imbalance of orders in interval B. For instance, at point $i$ (11:19:28), we observe a dramatic increase in the proportion of large inflow volume size, from 38.4\% to 54.5\%, indicating institutional intervention in the market and the initiation of upward price movement by the main force. At point $j$ (11:25:36), buyers have already established dominance as evidenced by a 33.0\% (54.2\% to 87.2\%) increase in the proportion of bid volume, and this is followed by a sustained upward trend in stock prices.
This illustrates that our model successfully captures the predictive signals preceding the upward price movement, which holds significant implications for investment decision-making.

\section{CONCLUSIONS}
In this paper, we introduce a method that utilizes order flow data to model and extract important trading details to enhance stock trend analysis. We propose a framework for extracting stock factors that leverages historical order flow data to capture contextual information and extract informative factors from the microstructure market. Moreover, we demonstrate the effectiveness of our method in two downstream tasks with varying temporal granularities by using a comprehensive year-long dataset from the real-world.

\bibliographystyle{ACM-Reference-Format}
\bibliography{cikm23}


\begin{thebibliography}{41}


\ifx \showCODEN    \undefined \def \showCODEN     #1{\unskip}     \fi
\ifx \showDOI      \undefined \def \showDOI       #1{#1}\fi
\ifx \showISBNx    \undefined \def \showISBNx     #1{\unskip}     \fi
\ifx \showISBNxiii \undefined \def \showISBNxiii  #1{\unskip}     \fi
\ifx \showISSN     \undefined \def \showISSN      #1{\unskip}     \fi
\ifx \showLCCN     \undefined \def \showLCCN      #1{\unskip}     \fi
\ifx \shownote     \undefined \def \shownote      #1{#1}          \fi
\ifx \showarticletitle \undefined \def \showarticletitle #1{#1}   \fi
\ifx \showURL      \undefined \def \showURL       {\relax}        \fi
\providecommand\bibfield[2]{#2}
\providecommand\bibinfo[2]{#2}
\providecommand\natexlab[1]{#1}
\providecommand\showeprint[2][]{arXiv:#2}

\bibitem[Avellaneda and Stoikov(2008)]%
        {avellaneda2008high}
\bibfield{author}{\bibinfo{person}{Marco Avellaneda} {and}
  \bibinfo{person}{Sasha Stoikov}.} \bibinfo{year}{2008}\natexlab{}.
\newblock \showarticletitle{High-frequency trading in a limit order book}.
\newblock \bibinfo{journal}{\emph{Quantitative Finance}} \bibinfo{volume}{8},
  \bibinfo{number}{3} (\bibinfo{year}{2008}), \bibinfo{pages}{217--224}.
\newblock


\bibitem[Barber and Odean(2013)]%
        {barber2013behavior}
\bibfield{author}{\bibinfo{person}{Brad~M Barber} {and}
  \bibinfo{person}{Terrance Odean}.} \bibinfo{year}{2013}\natexlab{}.
\newblock \showarticletitle{The behavior of individual investors}.
\newblock In \bibinfo{booktitle}{\emph{Handbook of the Economics of Finance}}.
  Vol.~\bibinfo{volume}{2}. \bibinfo{publisher}{Elsevier},
  \bibinfo{pages}{1533--1570}.
\newblock


\bibitem[Bousbaa et~al\mbox{.}(2023)]%
        {bousbaa2023financial}
\bibfield{author}{\bibinfo{person}{Zineb Bousbaa}, \bibinfo{person}{Javier
  Sanchez-Medina}, {and} \bibinfo{person}{Omar Bencharef}.}
  \bibinfo{year}{2023}\natexlab{}.
\newblock \showarticletitle{Financial Time Series Forecasting: A Data Stream
  Mining-Based System}.
\newblock \bibinfo{journal}{\emph{Electronics}} \bibinfo{volume}{12},
  \bibinfo{number}{9} (\bibinfo{year}{2023}), \bibinfo{pages}{2039}.
\newblock


\bibitem[Cavalcante et~al\mbox{.}(2016)]%
        {cavalcante2016computational}
\bibfield{author}{\bibinfo{person}{Rodolfo~C Cavalcante},
  \bibinfo{person}{Rodrigo~C Brasileiro}, \bibinfo{person}{Victor~LF Souza},
  \bibinfo{person}{Jarley~P Nobrega}, {and} \bibinfo{person}{Adriano~LI
  Oliveira}.} \bibinfo{year}{2016}\natexlab{}.
\newblock \showarticletitle{Computational intelligence and financial markets: A
  survey and future directions}.
\newblock \bibinfo{journal}{\emph{Expert Systems with Applications}}
  \bibinfo{volume}{55} (\bibinfo{year}{2016}), \bibinfo{pages}{194--211}.
\newblock


\bibitem[Chen et~al\mbox{.}(2019)]%
        {chen2019investment}
\bibfield{author}{\bibinfo{person}{Chi Chen}, \bibinfo{person}{Li Zhao},
  \bibinfo{person}{Jiang Bian}, \bibinfo{person}{Chunxiao Xing}, {and}
  \bibinfo{person}{Tie-Yan Liu}.} \bibinfo{year}{2019}\natexlab{}.
\newblock \showarticletitle{Investment behaviors can tell what inside:
  Exploring stock intrinsic properties for stock trend prediction}. In
  \bibinfo{booktitle}{\emph{Proceedings of the 25th ACM SIGKDD International
  Conference on Knowledge Discovery \& Data Mining}}.
  \bibinfo{pages}{2376--2384}.
\newblock


\bibitem[Chordia and Subrahmanyam(2004)]%
        {chordia2004order}
\bibfield{author}{\bibinfo{person}{Tarun Chordia} {and}
  \bibinfo{person}{Avanidhar Subrahmanyam}.} \bibinfo{year}{2004}\natexlab{}.
\newblock \showarticletitle{Order imbalance and individual stock returns:
  Theory and evidence}.
\newblock \bibinfo{journal}{\emph{Journal of Financial Economics}}
  \bibinfo{volume}{72}, \bibinfo{number}{3} (\bibinfo{year}{2004}),
  \bibinfo{pages}{485--518}.
\newblock


\bibitem[Ding et~al\mbox{.}(2020)]%
        {ding2020hierarchical}
\bibfield{author}{\bibinfo{person}{Qianggang Ding}, \bibinfo{person}{Sifan Wu},
  \bibinfo{person}{Hao Sun}, \bibinfo{person}{Jiadong Guo}, {and}
  \bibinfo{person}{Jian Guo}.} \bibinfo{year}{2020}\natexlab{}.
\newblock \showarticletitle{Hierarchical Multi-Scale Gaussian Transformer for
  Stock Movement Prediction.}. In \bibinfo{booktitle}{\emph{IJCAI}}.
  \bibinfo{pages}{4640--4646}.
\newblock


\bibitem[Fang et~al\mbox{.}(2021)]%
        {YuchenFang2021UniversalTF}
\bibfield{author}{\bibinfo{person}{Yuchen Fang}, \bibinfo{person}{Kan Ren},
  \bibinfo{person}{Weiqing Liu}, \bibinfo{person}{Dong Zhou},
  \bibinfo{person}{Weinan Zhang}, \bibinfo{person}{Jiang Bian},
  \bibinfo{person}{Yong Yu}, {and} \bibinfo{person}{Tie-Yan Liu}.}
  \bibinfo{year}{2021}\natexlab{}.
\newblock \showarticletitle{Universal Trading for Order Execution with Oracle
  Policy Distillation}.
\newblock \bibinfo{journal}{\emph{Research Papers in Economics}}
  (\bibinfo{year}{2021}).
\newblock


\bibitem[Feng et~al\mbox{.}(2018)]%
        {feng2018enhancing}
\bibfield{author}{\bibinfo{person}{Fuli Feng}, \bibinfo{person}{Huimin Chen},
  \bibinfo{person}{Xiangnan He}, \bibinfo{person}{Ji Ding},
  \bibinfo{person}{Maosong Sun}, {and} \bibinfo{person}{Tat-Seng Chua}.}
  \bibinfo{year}{2018}\natexlab{}.
\newblock \showarticletitle{Enhancing stock movement prediction with
  adversarial training}.
\newblock \bibinfo{journal}{\emph{arXiv preprint arXiv:1810.09936}}
  (\bibinfo{year}{2018}).
\newblock


\bibitem[Gould et~al\mbox{.}(2013)]%
        {gould2013limit}
\bibfield{author}{\bibinfo{person}{Martin~D Gould}, \bibinfo{person}{Mason~A
  Porter}, \bibinfo{person}{Stacy Williams}, \bibinfo{person}{Mark McDonald},
  \bibinfo{person}{Daniel~J Fenn}, {and} \bibinfo{person}{Sam~D Howison}.}
  \bibinfo{year}{2013}\natexlab{}.
\newblock \showarticletitle{Limit order books}.
\newblock \bibinfo{journal}{\emph{Quantitative Finance}} \bibinfo{volume}{13},
  \bibinfo{number}{11} (\bibinfo{year}{2013}), \bibinfo{pages}{1709--1742}.
\newblock


\bibitem[Guilbaud and Pham(2013)]%
        {guilbaud2013optimal}
\bibfield{author}{\bibinfo{person}{Fabien Guilbaud} {and}
  \bibinfo{person}{Huyen Pham}.} \bibinfo{year}{2013}\natexlab{}.
\newblock \showarticletitle{Optimal high-frequency trading with limit and
  market orders}.
\newblock \bibinfo{journal}{\emph{Quantitative Finance}} \bibinfo{volume}{13},
  \bibinfo{number}{1} (\bibinfo{year}{2013}), \bibinfo{pages}{79--94}.
\newblock


\bibitem[Hochreiter and Schmidhuber(1997)]%
        {hochreiter1997long}
\bibfield{author}{\bibinfo{person}{Sepp Hochreiter} {and}
  \bibinfo{person}{J{\"u}rgen Schmidhuber}.} \bibinfo{year}{1997}\natexlab{}.
\newblock \showarticletitle{Long short-term memory}.
\newblock \bibinfo{journal}{\emph{Neural computation}} \bibinfo{volume}{9},
  \bibinfo{number}{8} (\bibinfo{year}{1997}), \bibinfo{pages}{1735--1780}.
\newblock


\bibitem[Jegadeesh and Titman(2011)]%
        {jegadeesh2011momentum}
\bibfield{author}{\bibinfo{person}{Narasimhan Jegadeesh} {and}
  \bibinfo{person}{Sheridan Titman}.} \bibinfo{year}{2011}\natexlab{}.
\newblock \showarticletitle{Momentum}.
\newblock \bibinfo{journal}{\emph{Annu. Rev. Financ. Econ.}}
  \bibinfo{volume}{3}, \bibinfo{number}{1} (\bibinfo{year}{2011}),
  \bibinfo{pages}{493--509}.
\newblock


\bibitem[Lespagnol and Rouchier(2018)]%
        {lespagnol2018trading}
\bibfield{author}{\bibinfo{person}{Vivien Lespagnol} {and}
  \bibinfo{person}{Juliette Rouchier}.} \bibinfo{year}{2018}\natexlab{}.
\newblock \showarticletitle{Trading volume and price distortion: an agent-based
  model with heterogenous knowledge of fundamentals}.
\newblock \bibinfo{journal}{\emph{Computational Economics}}
  \bibinfo{volume}{51}, \bibinfo{number}{4} (\bibinfo{year}{2018}),
  \bibinfo{pages}{991--1020}.
\newblock


\bibitem[Li et~al\mbox{.}(2021)]%
        {li2021modeling}
\bibfield{author}{\bibinfo{person}{Wei Li}, \bibinfo{person}{Ruihan Bao},
  \bibinfo{person}{Keiko Harimoto}, \bibinfo{person}{Deli Chen},
  \bibinfo{person}{Jingjing Xu}, {and} \bibinfo{person}{Qi Su}.}
  \bibinfo{year}{2021}\natexlab{}.
\newblock \showarticletitle{Modeling the stock relation with graph network for
  overnight stock movement prediction}. In
  \bibinfo{booktitle}{\emph{Proceedings of the twenty-ninth international
  conference on international joint conferences on artificial intelligence}}.
  \bibinfo{pages}{4541--4547}.
\newblock


\bibitem[Liang et~al\mbox{.}(2021)]%
        {liang2021adaptive}
\bibfield{author}{\bibinfo{person}{Qianqiao Liang}, \bibinfo{person}{Mengying
  Zhu}, \bibinfo{person}{Xiaolin Zheng}, {and} \bibinfo{person}{Yan Wang}.}
  \bibinfo{year}{2021}\natexlab{}.
\newblock \showarticletitle{An Adaptive News-Driven Method for CVaR-sensitive
  Online Portfolio Selection in Non-Stationary Financial Markets.}. In
  \bibinfo{booktitle}{\emph{IJCAI}}. \bibinfo{pages}{2708--2715}.
\newblock


\bibitem[Liu et~al\mbox{.}(2020)]%
        {liu2020multi}
\bibfield{author}{\bibinfo{person}{Guang Liu}, \bibinfo{person}{Yuzhao Mao},
  \bibinfo{person}{Qi Sun}, \bibinfo{person}{Hailong Huang},
  \bibinfo{person}{Weiguo Gao}, \bibinfo{person}{Xuan Li},
  \bibinfo{person}{Jianping Shen}, \bibinfo{person}{Ruifan Li}, {and}
  \bibinfo{person}{Xiaojie Wang}.} \bibinfo{year}{2020}\natexlab{}.
\newblock \showarticletitle{Multi-scale Two-way Deep Neural Network for Stock
  Trend Prediction.}. In \bibinfo{booktitle}{\emph{IJCAI}}.
  \bibinfo{pages}{4555--4561}.
\newblock


\bibitem[Maglaras et~al\mbox{.}(2022)]%
        {maglaras2022deep}
\bibfield{author}{\bibinfo{person}{Costis Maglaras}, \bibinfo{person}{Ciamac~C
  Moallemi}, {and} \bibinfo{person}{Muye Wang}.}
  \bibinfo{year}{2022}\natexlab{}.
\newblock \showarticletitle{A deep learning approach to estimating fill
  probabilities in a limit order book}.
\newblock \bibinfo{journal}{\emph{Quantitative Finance}} \bibinfo{volume}{22},
  \bibinfo{number}{11} (\bibinfo{year}{2022}), \bibinfo{pages}{1989--2003}.
\newblock


\bibitem[Magris et~al\mbox{.}(2022)]%
        {magris2022bayesian}
\bibfield{author}{\bibinfo{person}{Martin Magris}, \bibinfo{person}{Mostafa
  Shabani}, {and} \bibinfo{person}{Alexandros Iosifidis}.}
  \bibinfo{year}{2022}\natexlab{}.
\newblock \showarticletitle{Bayesian Bilinear Neural Network for Predicting the
  Mid-price Dynamics in Limit-Order Book Markets}.
\newblock \bibinfo{journal}{\emph{arXiv preprint arXiv:2203.03613}}
  (\bibinfo{year}{2022}).
\newblock


\bibitem[McGroarty et~al\mbox{.}(2019)]%
        {mcgroarty2019high}
\bibfield{author}{\bibinfo{person}{Frank McGroarty}, \bibinfo{person}{Ash
  Booth}, \bibinfo{person}{Enrico Gerding}, {and} \bibinfo{person}{VL
  Chinthalapati}.} \bibinfo{year}{2019}\natexlab{}.
\newblock \showarticletitle{High frequency trading strategies, market fragility
  and price spikes: an agent based model perspective}.
\newblock \bibinfo{journal}{\emph{Annals of Operations Research}}
  \bibinfo{volume}{282}, \bibinfo{number}{1} (\bibinfo{year}{2019}),
  \bibinfo{pages}{217--244}.
\newblock


\bibitem[Moallemi and Wang(2022)]%
        {moallemi2022reinforcement}
\bibfield{author}{\bibinfo{person}{Ciamac~C Moallemi} {and}
  \bibinfo{person}{Muye Wang}.} \bibinfo{year}{2022}\natexlab{}.
\newblock \showarticletitle{A reinforcement learning approach to optimal
  execution}.
\newblock \bibinfo{journal}{\emph{Quantitative Finance}} \bibinfo{volume}{22},
  \bibinfo{number}{6} (\bibinfo{year}{2022}), \bibinfo{pages}{1051--1069}.
\newblock


\bibitem[Mukherjee et~al\mbox{.}(2023)]%
        {mukherjee2023stock}
\bibfield{author}{\bibinfo{person}{Somenath Mukherjee}, \bibinfo{person}{Bikash
  Sadhukhan}, \bibinfo{person}{Nairita Sarkar}, \bibinfo{person}{Debajyoti
  Roy}, {and} \bibinfo{person}{Soumil De}.} \bibinfo{year}{2023}\natexlab{}.
\newblock \showarticletitle{Stock market prediction using deep learning
  algorithms}.
\newblock \bibinfo{journal}{\emph{CAAI Transactions on Intelligence
  Technology}} \bibinfo{volume}{8}, \bibinfo{number}{1} (\bibinfo{year}{2023}),
  \bibinfo{pages}{82--94}.
\newblock


\bibitem[Nazareth and Reddy(2023)]%
        {nazareth2023financial}
\bibfield{author}{\bibinfo{person}{Noella Nazareth} {and}
  \bibinfo{person}{Yeruva Yenkata~Ramana Reddy}.}
  \bibinfo{year}{2023}\natexlab{}.
\newblock \showarticletitle{Financial applications of machine learning: a
  literature review}.
\newblock \bibinfo{journal}{\emph{Expert Systems with Applications}}
  (\bibinfo{year}{2023}), \bibinfo{pages}{119640}.
\newblock


\bibitem[Obizhaeva and Wang(2013)]%
        {obizhaeva2013optimal}
\bibfield{author}{\bibinfo{person}{Anna~A Obizhaeva} {and}
  \bibinfo{person}{Jiang Wang}.} \bibinfo{year}{2013}\natexlab{}.
\newblock \showarticletitle{Optimal trading strategy and supply/demand
  dynamics}.
\newblock \bibinfo{journal}{\emph{Journal of Financial Markets}}
  \bibinfo{volume}{16}, \bibinfo{number}{1} (\bibinfo{year}{2013}),
  \bibinfo{pages}{1--32}.
\newblock


\bibitem[Pereira et~al\mbox{.}(2020)]%
        {pereira2020trading}
\bibfield{author}{\bibinfo{person}{Gustavo Magno~Lopes Pereira},
  \bibinfo{person}{Eduardo Camilo-da Silva}, {and} \bibinfo{person}{Claudio
  Henrique da~Silveira Barbedo}.} \bibinfo{year}{2020}\natexlab{}.
\newblock \showarticletitle{Trading Imbalance, Liquidity and Stock Returns:
  Evidence from Brazil}.
\newblock \bibinfo{journal}{\emph{Latin American Business Review}}
  \bibinfo{volume}{21}, \bibinfo{number}{2} (\bibinfo{year}{2020}),
  \bibinfo{pages}{173--195}.
\newblock


\bibitem[Ro{\c{s}}u(2009)]%
        {rocsu2009dynamic}
\bibfield{author}{\bibinfo{person}{Ioanid Ro{\c{s}}u}.}
  \bibinfo{year}{2009}\natexlab{}.
\newblock \showarticletitle{A dynamic model of the limit order book}.
\newblock \bibinfo{journal}{\emph{The Review of Financial Studies}}
  \bibinfo{volume}{22}, \bibinfo{number}{11} (\bibinfo{year}{2009}),
  \bibinfo{pages}{4601--4641}.
\newblock


\bibitem[Ruff et~al\mbox{.}(2018)]%
        {ruff2018deep}
\bibfield{author}{\bibinfo{person}{Lukas Ruff}, \bibinfo{person}{Robert
  Vandermeulen}, \bibinfo{person}{Nico Goernitz}, \bibinfo{person}{Lucas
  Deecke}, \bibinfo{person}{Shoaib~Ahmed Siddiqui}, \bibinfo{person}{Alexander
  Binder}, \bibinfo{person}{Emmanuel M{\"u}ller}, {and} \bibinfo{person}{Marius
  Kloft}.} \bibinfo{year}{2018}\natexlab{}.
\newblock \showarticletitle{Deep one-class classification}. In
  \bibinfo{booktitle}{\emph{International conference on machine learning}}.
  PMLR, \bibinfo{pages}{4393--4402}.
\newblock


\bibitem[Schnaubelt(2022)]%
        {schnaubelt2022deep}
\bibfield{author}{\bibinfo{person}{Matthias Schnaubelt}.}
  \bibinfo{year}{2022}\natexlab{}.
\newblock \showarticletitle{Deep reinforcement learning for the optimal
  placement of cryptocurrency limit orders}.
\newblock \bibinfo{journal}{\emph{European Journal of Operational Research}}
  \bibinfo{volume}{296}, \bibinfo{number}{3} (\bibinfo{year}{2022}),
  \bibinfo{pages}{993--1006}.
\newblock


\bibitem[Shannon(1948)]%
        {shannon1948mathematical}
\bibfield{author}{\bibinfo{person}{Claude~E Shannon}.}
  \bibinfo{year}{1948}\natexlab{}.
\newblock \showarticletitle{A mathematical theory of communication}.
\newblock \bibinfo{journal}{\emph{The Bell system technical journal}}
  \bibinfo{volume}{27}, \bibinfo{number}{3} (\bibinfo{year}{1948}),
  \bibinfo{pages}{379--423}.
\newblock


\bibitem[Shen(2015)]%
        {shen2015order}
\bibfield{author}{\bibinfo{person}{Darryl Shen}.}
  \bibinfo{year}{2015}\natexlab{}.
\newblock \emph{\bibinfo{title}{Order imbalance based strategy in high
  frequency trading}}.
\newblock \bibinfo{thesistype}{Ph.\,D. Dissertation}. \bibinfo{school}{oxford
  university}.
\newblock


\bibitem[Shi and Cartlidge(2022)]%
        {shi2022state}
\bibfield{author}{\bibinfo{person}{Zijian Shi} {and} \bibinfo{person}{John
  Cartlidge}.} \bibinfo{year}{2022}\natexlab{}.
\newblock \showarticletitle{State Dependent Parallel Neural Hawkes Process for
  Limit Order Book Event Stream Prediction and Simulation}. In
  \bibinfo{booktitle}{\emph{Proceedings of the 28th ACM SIGKDD Conference on
  Knowledge Discovery and Data Mining}}. \bibinfo{pages}{1607--1615}.
\newblock


\bibitem[Shi et~al\mbox{.}(2021)]%
        {shi2021lob}
\bibfield{author}{\bibinfo{person}{Zijian Shi}, \bibinfo{person}{Yu Chen},
  {and} \bibinfo{person}{John Cartlidge}.} \bibinfo{year}{2021}\natexlab{}.
\newblock \showarticletitle{The LOB recreation model: Predicting the limit
  order book from TAQ history using an ordinary differential equation recurrent
  neural network}. In \bibinfo{booktitle}{\emph{Proceedings of the AAAI
  Conference on Artificial Intelligence}}, Vol.~\bibinfo{volume}{35}.
  \bibinfo{pages}{548--556}.
\newblock


\bibitem[Sirignano(2019)]%
        {sirignano2019deep}
\bibfield{author}{\bibinfo{person}{Justin~A Sirignano}.}
  \bibinfo{year}{2019}\natexlab{}.
\newblock \showarticletitle{Deep learning for limit order books}.
\newblock \bibinfo{journal}{\emph{Quantitative Finance}} \bibinfo{volume}{19},
  \bibinfo{number}{4} (\bibinfo{year}{2019}), \bibinfo{pages}{549--570}.
\newblock


\bibitem[Thakkar and Chaudhari(2021)]%
        {thakkar2021fusion}
\bibfield{author}{\bibinfo{person}{Ankit Thakkar} {and} \bibinfo{person}{Kinjal
  Chaudhari}.} \bibinfo{year}{2021}\natexlab{}.
\newblock \showarticletitle{Fusion in stock market prediction: a decade survey
  on the necessity, recent developments, and potential future directions}.
\newblock \bibinfo{journal}{\emph{Information Fusion}}  \bibinfo{volume}{65}
  (\bibinfo{year}{2021}), \bibinfo{pages}{95--107}.
\newblock


\bibitem[Tsantekidis et~al\mbox{.}(2017)]%
        {tsantekidis2017forecasting}
\bibfield{author}{\bibinfo{person}{Avraam Tsantekidis},
  \bibinfo{person}{Nikolaos Passalis}, \bibinfo{person}{Anastasios Tefas},
  \bibinfo{person}{Juho Kanniainen}, \bibinfo{person}{Moncef Gabbouj}, {and}
  \bibinfo{person}{Alexandros Iosifidis}.} \bibinfo{year}{2017}\natexlab{}.
\newblock \showarticletitle{Forecasting stock prices from the limit order book
  using convolutional neural networks}. In \bibinfo{booktitle}{\emph{2017 IEEE
  19th conference on business informatics (CBI)}}, Vol.~\bibinfo{volume}{1}.
  IEEE, \bibinfo{pages}{7--12}.
\newblock


\bibitem[Wang et~al\mbox{.}(2021)]%
        {wang2021hierarchical}
\bibfield{author}{\bibinfo{person}{Heyuan Wang}, \bibinfo{person}{Shun Li},
  \bibinfo{person}{Tengjiao Wang}, {and} \bibinfo{person}{Jiayi Zheng}.}
  \bibinfo{year}{2021}\natexlab{}.
\newblock \showarticletitle{Hierarchical Adaptive Temporal-Relational Modeling
  for Stock Trend Prediction.}. In \bibinfo{booktitle}{\emph{IJCAI}}.
  \bibinfo{pages}{3691--3698}.
\newblock


\bibitem[Wang et~al\mbox{.}({[n.\,d.]})]%
        {wangheterogeneous}
\bibfield{author}{\bibinfo{person}{Heyuan Wang}, \bibinfo{person}{Tengjiao
  Wang}, \bibinfo{person}{Shun Li}, \bibinfo{person}{Shijie Guan},
  \bibinfo{person}{Jiayi Zheng}, {and} \bibinfo{person}{Wei Chen}.}
  \bibinfo{year}{[n.\,d.]}\natexlab{}.
\newblock \showarticletitle{Heterogeneous Interactive Snapshot Network for
  Review-Enhanced Stock Profiling and Recommendation}.
\newblock  (\bibinfo{year}{[n.\,d.]}).
\newblock


\bibitem[Wang et~al\mbox{.}(2022)]%
        {wang2022adaptive}
\bibfield{author}{\bibinfo{person}{Heyuan Wang}, \bibinfo{person}{Tengjiao
  Wang}, \bibinfo{person}{Shun Li}, \bibinfo{person}{Jiayi Zheng},
  \bibinfo{person}{Shijie Guan}, {and} \bibinfo{person}{Wei Chen}.}
  \bibinfo{year}{2022}\natexlab{}.
\newblock \showarticletitle{Adaptive Long-Short Pattern Transformer for Stock
  Investment Selection}. IJCAI.
\newblock


\bibitem[Wei et~al\mbox{.}(2019)]%
        {wei2019model}
\bibfield{author}{\bibinfo{person}{Haoran Wei}, \bibinfo{person}{Yuanbo Wang},
  \bibinfo{person}{Lidia Mangu}, {and} \bibinfo{person}{Keith Decker}.}
  \bibinfo{year}{2019}\natexlab{}.
\newblock \showarticletitle{Model-based reinforcement learning for predictions
  and control for limit order books}.
\newblock \bibinfo{journal}{\emph{arXiv preprint arXiv:1910.03743}}
  (\bibinfo{year}{2019}).
\newblock


\bibitem[Xu et~al\mbox{.}(2022)]%
        {xu2022multi}
\bibfield{author}{\bibinfo{person}{Borui Xu}, \bibinfo{person}{Tong Zhang},
  {and} \bibinfo{person}{Weiguo Liu}.} \bibinfo{year}{2022}\natexlab{}.
\newblock \showarticletitle{A Multi-scale Convolution and Gated Recurrent Unit
  Based Network for Limit Order Book Prediction}. In
  \bibinfo{booktitle}{\emph{International Conference on Knowledge Science,
  Engineering and Management}}. Springer, \bibinfo{pages}{71--84}.
\newblock


\bibitem[Zhang et~al\mbox{.}(2019)]%
        {zhang2019deeplob}
\bibfield{author}{\bibinfo{person}{Zihao Zhang}, \bibinfo{person}{Stefan
  Zohren}, {and} \bibinfo{person}{Stephen Roberts}.}
  \bibinfo{year}{2019}\natexlab{}.
\newblock \showarticletitle{Deeplob: Deep convolutional neural networks for
  limit order books}.
\newblock \bibinfo{journal}{\emph{IEEE Transactions on Signal Processing}}
  \bibinfo{volume}{67}, \bibinfo{number}{11} (\bibinfo{year}{2019}),
  \bibinfo{pages}{3001--3012}.
\newblock


\end{thebibliography}

\appendix

\end{document}